\documentclass[]{iopart}

\usepackage{graphicx}
\usepackage{epsfig}
\usepackage{subfigure}
\usepackage{sidecap}
\usepackage{iopams}
\usepackage{hyperref}

\begin{document}

\title[Nonlinear Resonances in Bose-Einstein Condensate]{Controlled Generation of Nonlinear Resonances through Sinusoidal Lattice Modes in Bose-Einstein Condensate}

\author{Priyam Das}
  \address{Institute of Nuclear Science, Hacettepe University, Ankara - 06800, Turkey}
\ead{daspriyam3@gmail.com}

\author{Prasanta K. Panigrahi}
 \address{Indian Institute of Science Education and Research - Kolkata, Mohanpur - 741246, India}



\begin{abstract}
We study Bose-Einstein condensate in the combined presence of time modulated optical lattice and harmonic trap in the mean-field approach. Through the self-similar method, we show the existence of sinusoidal lattice modes in this inhomogeneous system, commensurate with the lattice potential. A significant advantage of this system is wide tunability of the parameters through chirp management. The combined effect of the interaction, harmonic trap and lattice potential leads to the generation of nonlinear resonances, exactly where the matter wave changes its direction. When the harmonic trap is switched off, the BEC undergoes a nonlinear compression for the static optical lattice potential. For better understanding of chirp management and the nature of the sinusoidal excitation, we investigate the energy spectrum of the condensate, which clearly reveals the generation of nonlinear resonances in the appropriate regime. We have also identified a classical dynamical phase transition occurring in the system, where loss of superfluidity takes the superfluid phase to an insulating state.
\end{abstract}

\indent\pacs{03.75.Kk, 67.85.De, 67.85.Hj}

{\it Keywords}: Bose-Einstein condensate, Optical lattice, Nonlinear resonance

{\let\newpage\relax\maketitle}






\section{Introduction}

During the last two decades, the area of Bose-Einstein condensate (BEC) has seen significant advances, both in theoretical and experimental fronts \cite{dalfovo1999,leggett2001}. Response of the BEC changes drastically, when the system is loaded in an optical lattice (OL) potential, which is an area of active research  ~\cite {bloch2005ultracold,bloch2005quantum,jaksch2005cold,morsch2006dynamics,bloch2008,lin2008}. The observed behavior of the system can be divided into two parts: static and dynamic cases. In the former one, for the case of deep optical lattices, a metal to insulator quantum phase transition \cite{greiner2002quantum}, driven by the quantum fluctuations can occur, whereas, for shallow lattices, the same can occur through a classical dynamical phase transition \cite{smerzi2002dynamical,cataliotti2003superfluid,adhikari2003dynamical,das2009loss}, driven by modulational instability. The dynamic case is particularly interesting as it leads to a number of interesting phenomena in quantum and classical domains. These include, the observation of dynamical localization \cite{klappauf1998observation, ringot2000experimental}, Anderson metal-insulator tunneling in momentum space \cite{chabe2008}, gapped modes in degenerate quantum gases \cite{jordens2008,huber2008}, realization of nonlinear coherent modes by means of resonant excitations \cite{yukalov1997,yukalov2001,yukalov2002} etc. Optical lattices can be made time dependent by temporal variations of the depth and the width of the lattice \cite{fallani2004}. The lattice can be moved at a velocity $v = (\lambda/2)\delta\nu$, where $\lambda/2$ is the spatial period of the potential and $\delta \nu$ is the stable detuning between the two laser beams, which also allows the lattice to accelerate. The amplitude of the lattice potential can be tuned through the following relation: $V_{0}\approx 2 \hbar \Omega_{R}/E_{R}$, $\Omega_{R}$ being the Rabi frequency and $E_{R} = 2 \hbar^{2} \pi^{2}/ m \lambda^{2}$, the recoil energy.

A number of recent studies have investigated the condensate dynamics in the presence of time modulated optical lattices. It is worth pointing out in this context that, Fallani {\it et. al.}, have observed dynamical instability in one dimensional moving optical lattice \cite{fallani2004}. Recently, the region of the dynamical instability of the condensate with higher order nonlinearities has been explored in a time dependent lattice potential  \cite{wamba2014dynamical}. A detailed analysis has been made in \cite{pederson2013}, for the creation, evolution and manipulation of wave packets in the combined potential of optical lattice and a harmonic oscillator (HO) trap. Some time back, Madison et. al. \cite{madison1997quantum}, studied the quantum transport of atoms in the presence of an accelerating lattice. They observed resonances, when a small oscillatory trap was added to the system. The number of atoms decreases rapidly when the resonances occur. The spatio-temporal dynamics of the condensate has also been investigated in the presence of moving optical lattice, using direct perturbation and Melnikov-function methods \cite{li2007}. The existence of sub-diffractive solitons and chaos has been shown to exist for a time dependent shallow optical lattice  \cite{staliunas2008}. These authors also showed the possibility of suppression of chaos and its control through the intensity of the lattice potential. Recently, nonlinear resonance phenomenon has been studied in BEC, where the occurrence of resonances has been investigated through periodic modulation of the scattering length \cite{s2014,verma2012oscillations,ivana2013}. Unlike the methods opted in these studies, we take a different approach by introducing a lattice potential and show that it is possible to generate the nonlinear resonances in BEC, through controlled chirp management.
\par

In this paper, we present a detailed study of the dynamics of BEC in the combined presence of a general time dependent optical lattice and a harmonic oscillator trap, for both attractive and repulsive atom-atom interactions. The exact sinusoidal solutions are naturally chirped, which allows them to be compressed and accelerated in a controlled manner.  The presence of the trap drastically changes the dynamics of the condensate. We show the manifestation of nonlinear resonances, which arise due to the combined effect of harmonic trap, interactions and lattice potential. These resonances occur, when the matter wave changes its direction. The density of the condensate is found to be maximum at this point. In a harmonic trap, the center of mass (COM) undergoes periodic oscillation at trap frequency, as a consequence of the Kohn theorem, which makes the matter wave change directions periodically. We have shown that by engineering the trap parameter, one can have various COM profiles that lead to interesting physics. We then consider the BEC in the presence of optical lattice, when the harmonic trap is switched off. In case of a static OL potential, the BEC undergoes a rapid nonlinear compression, leading to a resonance behavior. This phenomenon of rapid nonlinear compression is analogous to the effective pulse compression in nonlinear optical fiber \cite{moores1996}. For the sake of completeness, we also analyze the system in the presence of an expulsive harmonic trap, where BEC spreads out. For gaining a deeper insight of these resonances, we investigate the energy spectrum of BEC both analytically and numerically. In the combined presence of harmonic trap and OL potential, the condensate shows a resonant increase in energy at certain points in the scaled time variable, which supports our observation through the density distribution. In case of static OL, a sudden increase of energy mimics the resonance behavior of the system. For comparison, we perform a numerical simulation, which clearly  shows the generation of nonlinear resonances, similar to the analytically obtained results. This system also exhibits a classical dynamical superfluid insulator transition (DSIT), where superfluidity breaks down and the condensate transits to an insulating phase. This phase transition is of the first order and occurs due to modulational instability \cite{smerzi2002dynamical}. We show the stability of the obtained solutions in the repulsive domain, using Vakitov-Kolokolov (VK) criterion \cite{vakhitov1974}.

The paper is organized in the following way: in Sec. 2, we present the theoretical model and derive the governing mean-field equation that describes the system. The exact solutions, described in Sec. 3, are found in different regions, describing the sinusoidal excitations in the presence of both lattice and harmonic trap. In Sec. 4, the nature of these obtained solutions and generation of nonlinear resonances in different parameter regimes have been analyzed. The generation of nonlinear resonances is discussed in Secs. 4.1 and 4.2. The energy spectrum has been investigated in Sec. 5, where we find that it supports the generation of the nonlinear resonances. Analytical, as well as numerical results, are illustrated respectively, in Secs. 5.1 and 5.2. The occurrence of dynamical phase transition in the system is described in Sec. 6. This classical phase transition occurs at the point, where the energy expression becomes non-analytic. In Sec. 7, we perform stability analysis using Vakhitov-Kolokolov criterion and show that the obtained solutions are stable for repulsive atom-atom interaction. We draw our conclusion in Sec. 8 and discuss the possible future prospects.


\section{Theoretical model}

We consider here a  BEC immersed in a time modulated OL potential and highly elongated harmonic trap in the longitudinal direction. In the ultra-cold regime, at very low temperature, the mean field Gross-Pitaevskii (GP) equation well captures the condensate dynamics:
\begin{eqnarray}
i \hbar \frac{\partial \Psi}{\partial t} = \left(- \frac{\hbar^{2}}{2 m} \nabla^{2} + V_{ext}(r, t) + U(t) |\Psi|^{2} - \bar{\nu}(t) \right)\Psi.
\end{eqnarray}
Here, $m$ is the mass of the atoms, $V_{ext}(r, t) = V_{tr}(x,y) + V_{l}(z,t)$ is the external trapping potential; $V_{tr}(x,y)$ and $V_{l}(z,t)$, respectively, represent the potential along the transverse and longitudinal directions. $U(t) = \frac{4 \pi \hbar^{2} a_{s}(t)}{m}$ is the
strength of the atom-atom interaction and $\bar{\nu}(t)$ is the time
dependent chemical potential. For obtaining a cigar-shaped BEC, one applies a
strong oscillator trapping potential of frequency $\omega_{\perp}$,
$V_{tr}(x,y) = \frac{1}{2}m \omega^{2}_{\perp} (x^{2} + y^{2})$, along the transverse direction.
Assuming tight transverse confinement, the trial wave function can be factorized:
 $\Psi = \psi(z,t) \phi_{0}(x,y)$, where $\phi_{0}(x,y)$ is represented
by a Gaussian ansatz \cite{salasnich2002effective,salasnich2007matter}:
\begin{eqnarray}
\phi_{0}(x,y) = \sqrt{\frac{1}{\pi a^{2}_{\perp}}} e^{- \frac{x^{2} +
    y^{2}}{2 a^{2}_{\perp}}},
\end{eqnarray}
with $a_{\perp} = \sqrt{\frac{\hbar}{m \omega_{\perp}}}$. From now onwards for convenience, we take $\psi(x,y) \rightarrow \psi$ throughout this paper. In the weak coupling regime, one finds that the quasi one dimensional GP equation takes the following form in terms of the dimensionless variable \cite{choi1999bose,absullaev2005gap,utpal2010complex}:
\begin{eqnarray}
i \frac{\partial \psi}{\partial t} = - \frac{1}{2} \frac{\partial^{2} \psi}{\partial z^{2}} + V_{l}(z,t) \psi + g(t)
|\psi|^{2} \psi - \nu(t) \psi . \label{1D-NLSE}
\end{eqnarray}
Here, $V_{l}(z,t)$ is the trapping potential along the longitudinal direction.
The scaled two-body interaction is given by, $g(t) = \frac{m \omega_{\perp}}{2 \pi \hbar} (\frac{m}{\hbar^{2} k_{R}}) U(t)$, where, $k_{R}$ is the wave vector. The dimensionless time, spatial coordinate and the wave function
are, respectively, scaled as, $t' \rightarrow (\hbar k_{R}^{2}/m) t$, $z'
\rightarrow k_{R} z$ and $\psi'(z,t) \rightarrow \psi(z,t)/\sqrt{k_{R}}$ \cite{choi1999bose} (for notational convenience, we replace $t'$ by $t$, $z'$ by $z$ and $\psi'$ by $\psi$ in Eq. (\ref{1D-NLSE})). The chemical
potential $\bar{\nu}(t)$ has been scaled in terms of the recoil energy $E_{R} = \frac{\hbar^{2}
  k_{R}^{2}}{2 m}$, such that $\bar{\nu}(t) = E_{R} \nu(t)$.

Keeping in mind the fact that most of the atomic systems are inhomogeneous due to the presence of the magnetic or optical traps, we consider the BEC in presence of the harmonic oscillator potential, in addition to the optical lattice: $V_{l}(z,t) = V_{0}(t) \cos^{2}(T(z,t)) + \frac{1}{2}M(t) z^{2}$, with $T(z,t) = k(t)(z - l(t))$. For notational convenience, we use $T(z,t)$ as $T$, throughout this paper. Here, $l(t) = \int^{t}_{0}v(t')dt'$ is the position, with $v(t)$ being the COM motion of the condensate and $k(t)$ is the inverse of the width. Since, the potential $V_{l}(z,t)$ is scaled by the recoil energy $E_{R}$, the time dependent amplitude of the lattice potential $V_{0}(t)$ and frequency of the harmonic trap $M(t)$, are both scaled by the recoil energy $E_{R}$ ($V_{0}(t) = V'_{0}(t) E_{R}$ and $M(t) = M'(t) E_{R}$, and henceforth, we replace $V'_{0}(t) \rightarrow V_{0}(t)$, $M'(t) \rightarrow M(t)$ in the expression of $V_{l}(z,t)$ above). Control of BEC and its solitonic and sinusoidal excitations through the time dependent trap parameter $M(t)$ is an area of significant current interest \cite{serkin2000novel,kruglov2003exact,atre2006class,das2009sinusoidal,kolovsky2010bose}. It has been observed that, the center of mass of solitons, as well as its width and amplitude get coupled to the trap parameters, through which they can be accelerated or compressed \cite{ranjani2008soliton,ranjani2010construction}, which is useful for coherent atom optics \cite{lenz1993,busch2002,keterle2002}.

\section{Sinusoidal excitations}

A number of methods have been employed in order to solve these type of systems. Among them, variational approximation method and perturbation methods are well known. In this paper, we follow a different approach known as self-similar method, to examine the exact analytical solutions and corresponding dynamics of this system with harmonic trap and lattice potentials. The first step of this method is to employ an ansatz solution, which consists of a non-trivial phase that is necessarily chirped. This type of chirped phase often arises in nonlinear optics and fiber optics as acceleration induced inhomogeneity, owing its origin to the interplay between the interaction and harmonic trap. We, therefore assume the following ansatz for the BEC profile \cite{das2009loss,atre2006class,das2009sinusoidal}:
\begin{eqnarray}
\psi(z,t) = \sqrt{k(t) \sigma(T)} e^{i [\chi(T) + \phi(z,t)]}.
\end{eqnarray}
where, $\sigma(T)$ is the density and $\chi(T)$ is the nontrivial phase, controlling the supercurrent of the condensate.
The chirped phase $\phi(z,t)$, is of the form: $\phi(z,t) = -\frac{1}{2} c(t) z^{2}$.
Current conservation yields, $\frac{\partial \chi(T)}{\partial T} = \frac{2 \delta}{\sigma(T)}$, where $\delta = \frac{1}{4 \kappa}\sqrt{(1 -   2 \mu + 2 \alpha)(1 - 2 \mu)}$. Here, the two-body interaction strength $g(t)$, the lattice depth $V_{0}(t)$ and the chemical potential $\nu(t)$ are rescaled by the inverse of the width of the condensate $k(t)$, such as, $g(t) = \kappa k(t)$, $V_{0}(t) = \alpha k^{2}(t)$ and $\nu(t) = \mu k^{2}(t)$, respectively. $\kappa$, $\alpha$ and $\mu$ are, respectively, the reduced two-body interaction strength, amplitude of the lattice potential and chemical potential.
The chirp parameter $c(t)$ can be found from the following Riccati equation:
\begin{equation}
\frac{\partial c(t)}{\partial t} - c^{2}(t) = M(t).
\label{ricatti}
\end{equation}
The inverse of width $k(t)$ is related to the chirp parameter $c(t)$ by, $\frac{\partial k(t)}{\partial t} =  k(t)c(t)$. From Kohn's theorem \cite{kohn1961}, it is known that for a BEC, confined in a harmonic trap with constant frequency, the COM oscillates with the frequency of the trap, giving rise to a resonance known as the Kohn mode. It is also known that when the trap frequency varies with time, the COM motion is no longer decoupled from the trap \cite{japha2002motion}. In order to have a complete control over the spatio-temporal dynamics of these sinusoidal excitations, we separate the equation governing the Kohn mode \cite{atre2006class}: $\frac{\partial l(t)}{\partial t} + c(t) l(t) = k(t) u$. Here, we have introduced a parameter $u$, which is the velocity of the condensate in the absence of the trap. When the harmonic trap is switched off ($M(t) = 0$), the COM moves with a constant velocity $u$, as is discussed in Sec. 4.4. The coupling of the center of mass motion with other degrees of freedom and trap parameters, is worth investigating when both trap and lattice co-exist. Interestingly, we observe that the above can also be written in terms of the harmonic trap, from where the consequences of the Kohn's theorem can be understood directly:
\begin{eqnarray}
\frac{\partial^{2} l(t)}{\partial t^{2}} + M(t) l(t) = 0.
\label{comp}
\end{eqnarray}
Therefore, it can be clearly seen that the COM solely depends on the harmonic trap. Now, the real part of the GP equation, in terms of the density, reduces to \cite{das2009loss,atre2006class,das2009sinusoidal,carr2001bose}:
\begin{eqnarray}
- \frac{1}{4} \sigma(T) \frac{\partial^{2} \sigma(T)}{\partial T^{2}} &-&
\frac{1}{8} \left(\frac{\partial \sigma(T)}{\partial T}\right)^{2}
- \mu \sigma^{2}(T) + \kappa \sigma^{3}(T) \nonumber \\ &+& \alpha \cos^{2}(T) \sigma^{2}(T)
+ 2 \delta^{2} = 0.
\end{eqnarray}
The first two terms correspond to the contribution from the dispersion term in Eq. (\ref{1D-NLSE}); third, fourth and fifth terms, respectively address to the chemical potential, atom-atom interactions and the lattice potential. The last term originates from the current conservation and is directly related to the supercurrent of the condensate. For constant $\alpha$, one finds a self-similar solution in terms of the density:
\begin{equation}
\sigma(T) = a + b \cos^{2} (T). \label{soln}
\end{equation}
where, $a = \frac{2 \mu - 1}{2 \kappa}$ and $b = - \frac{\alpha}{\kappa}$. The periodicity of the density modulation is same as that of the lattice potential as seen in Eq. \ref{soln}. Notice that, for $\sigma (T)$ to be positive semi-definite, considering the atom-atom interaction to be repulsive, $\mu \geq \alpha + 1/2$, when $\alpha$ is positive. For, negative $\alpha$, $\mu \geq 1/2$. When the interaction is taken to be attractive, for positive and negative $\alpha$, $\mu \geq 1/2$ and $\mu \geq \alpha + 1/2$, respectively.

\begin{figure}[t]
\begin{center}
\includegraphics[scale=0.5]{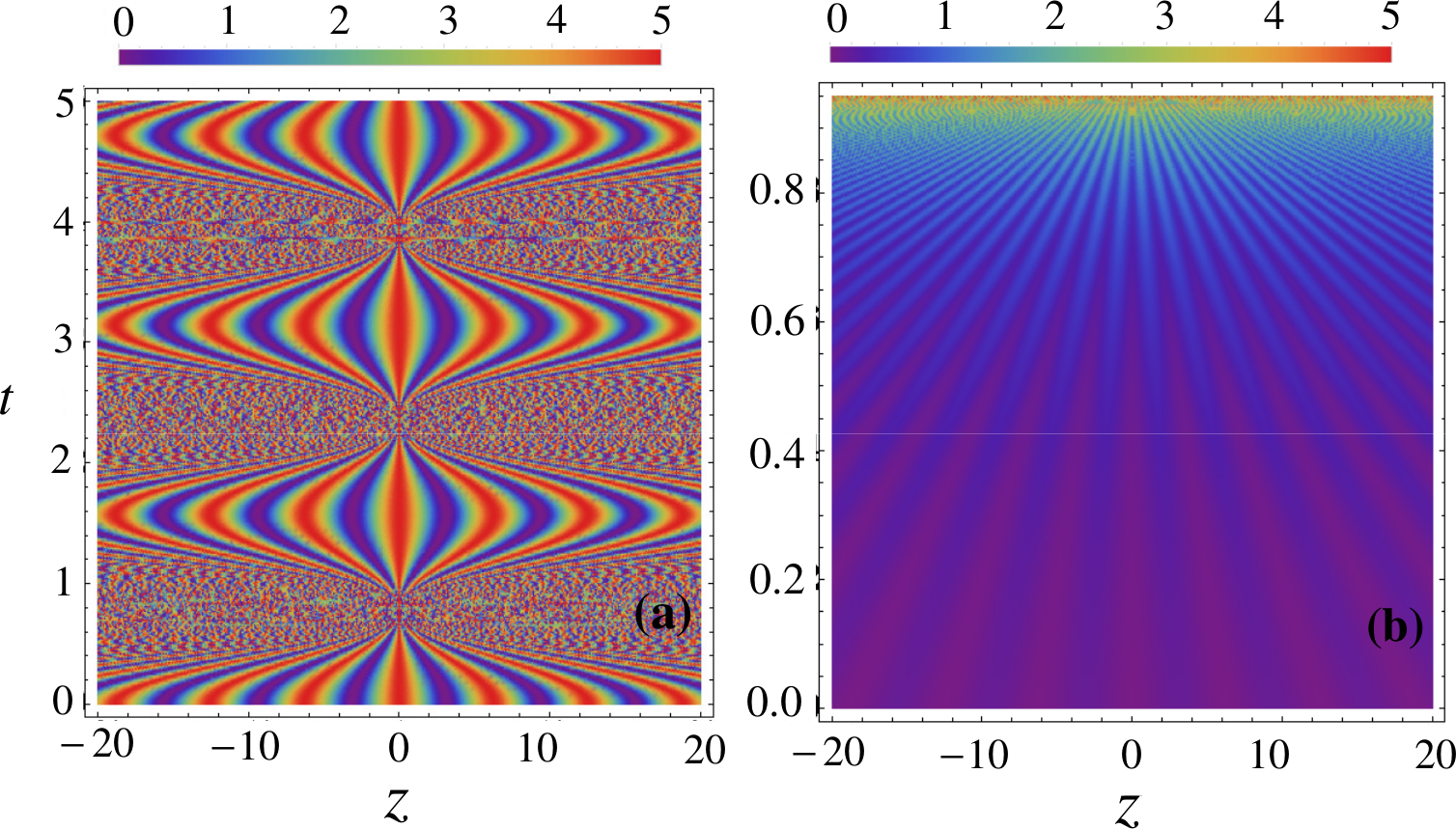}
\caption{Fig.(a) represents the density in presence of a regular harmonic trap, where at the point of nonlinear
         compression, the matter wave changes its direction. A sharp resonance behavior is observed, when the nonlinear compression takes
         place. Fig.(b) corresponds to the expulsive oscillator, where the
         BEC spreads.  Here, the time $t$ and space co-ordinate $z$ have been represented as dimensionless parameters after
         rescaling. In these figures, the parameter values that have been used
         are: $c_{0} = 1, k_{0} = 0.5, l_{0} = 0.4, \alpha = 0.15, u = 0.8, \kappa = 0.8$ and $q = 1$.}
\label{density}
\end{center}
\end{figure}

\section{Results and discussions}

Resonances, in general, occur when the frequency of the external driving force matches with the natural frequency of the system. Resonances are well understood in various linear dynamical systems. The investigations of such resonances in nonlinear systems are far more exciting, as well as nontrivial. The presence of nonlinearity in the system drastically changes the dynamics of such resonances. In this section, we analyze the obtained solutions in different parameter regimes and show the generation of nonlinear resonances through the time modulation of OL, both in presence and absence of harmonic trap, with controlled chirp management.

\subsection{Regular harmonic trap: $M(t) = q^{2}$}

\begin{figure}[h]
\begin{center}
\includegraphics[scale=0.5]{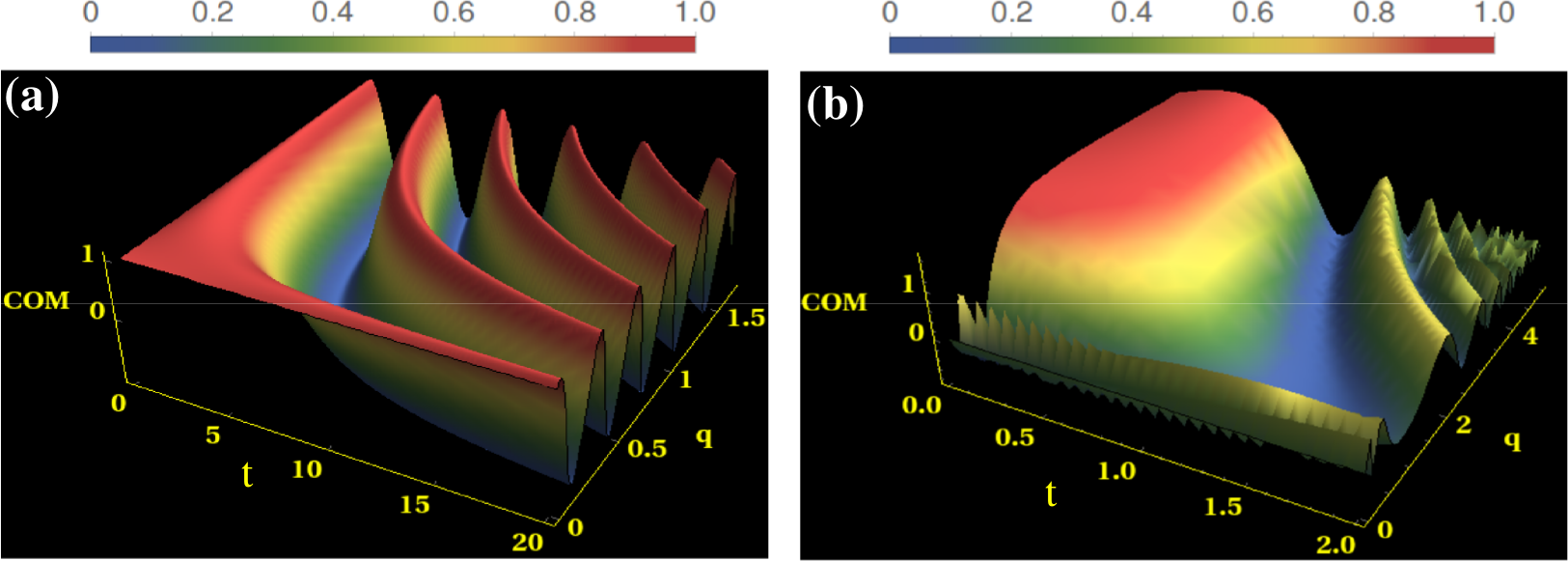}
\caption{COM profile is depicted in Fig.(a), in presence of a regular harmonic trap. It shows that the COM oscillates with the trap frequency. Fig.(b) presents the COM profile, described by the Bessel functions, when the trap parameter varies exponentially. The parameter values used here are same as in Fig.(\ref{density}).}
\label{com}
\end{center}
\end{figure}

  In the first case, we assume the trapping parameter as constant: $M(t) = q^{2}$, which represents a
  regular harmonic oscillator trap. Here, two solutions are possible: $(i)$ $c(t) = q \tan
  q t$ and $(ii)$ $c(t) = - q \cot q t$. Consequently, the other
  parameters can be obtained. For $(i)$, $k(t) = k_{0}/ \cos q t$ and
  $V_{0}(t) = V_{0}/ \cos^{2} q t$ and for $(ii)$, $k(t) =k_{0}/\sin q
  t$ and $V_{0}(t) = V_{0}/ \sin^{2} q t$, with $V_{0} = \alpha
  k^{2}_{0}$. The position of the matter wave is oscillatory: $(i)$
  $l(t) = l_{0} \cos q t$ and $(ii)$ $l(t) = l_{0} \sin q t$,
  respectively, with the initial position $l_{0}$. It is clearly seen that
  the Kohn mode oscillates with the trap frequency $q$, as predicted by
  Kohn's theorem. The condensate eventually evolves into periodic
  oscillations under the modulation of optical lattice and harmonic
  trap. As depicted in Fig.(\ref{density}a), the sinusoidal wave propagates with a
  periodicity $\pi$ and at the point of nonlinear compression, the
  density becomes maximum, wherein, the matter wave reverses
  its direction.  Here, the supercurrent takes the form,
\begin{equation}
J(t) = k(t) |\psi(z,t)|^{2} \frac{\partial \chi(T)}{\partial T} = 2 \delta k^{2}(t).
\label{sc}
\end{equation}
Fig.(\ref{com}), shows the center of mass profile with respect to time and trap frequency. In this case,  we consider a general solution of Eq.(\ref{comp}).  In this regard, it is worth mentioning that one can engineer the trap parameter on demand, in order to make the COM profile of a desired form. As an example, we have explicitly checked that if the trap parameter varies exponentially ($M(t) = e^{q t}$), the COM profile is described by Bessel functions of both first and second kind, which are solutions of Eq. (\ref{comp}). The profile is shown in Fig.(\ref{com}b), which shows that the COM no longer oscillates with the trap frequency and shows damped oscillations.  One can investigate in detail the effect of the trap parameter on COM, which we have left for future investigation.

\subsection{Static lattice: $M(t) = 0$ and $l(t) = 0$}

We now consider the case of a static optical lattice potential by putting $l(t) = 0$. For effective pulse compression  \cite{moores1996},  we consider a time dependent chirp parameter of the form, $c(t) = \frac{c_{0}}{1 - c_{0} t}$, and obtain, $k(t) = \frac{k_{0}}{1 - c_{0} t}$, with the initial condition $c(t=0) = c_{0}$ and $k(t=0) = k_{0}$. The compression of the matter wave is shown in Fig.(\ref{densitycd}b). Chirping leads to a rapid compression of the BEC profile and the density tends to a singularity, when $t \rightarrow 1/c_{0}$. It is evident that these self-similar excitations are purely controlled through the chirped phase, under certain parametric conditions, which in turn leads to efficient compression and amplification. The rapid nonlinear compression shown here is analogous to the findings by Moores in optical fiber \cite{moores1996}.


\subsection{Expulsive oscillator:$M(t) = -q^{2}$}

  For the sake of completeness, we consider $M(t) = -q^{2}$, which corresponds to an
  expulsive oscillator. The Riccati equation in this case, takes the form: $\frac{\partial c(t)}{\partial t}
  = c^{2}(t) - q^{2}$. Similar to the previous case, we have two
  solutions: $(i)$ $c(t) = - q \tanh q t$ and $(ii)$ $c(t) = - q \coth
  q t$.  The inverse of width and lattice amplitude for $(i)$ are:
  $k(t) = k_{0}/ \cosh q t$, $V_{0}(t) = V_{0}/ \cosh^{2} q t$ and for
  $(ii)$ $k(t) = k_{0}/ \sinh q t$, $V_{0}(t) = V_{0}/ \sinh^{2} q t$.
  The density profile for this case is shown in Fig. (\ref{densitycd}a). As time
  increases, the amplitude of the propagating wave decreases, making
  the BEC spread out. The distortion in the density profile, as seen in Fig.(\ref{densitycd}a),
  is due to the presence of the trap parameter.


\subsection{Moving lattice:  $M(t) = 0$}

\begin{figure}[t]
\begin{center}
\includegraphics[scale=0.5]{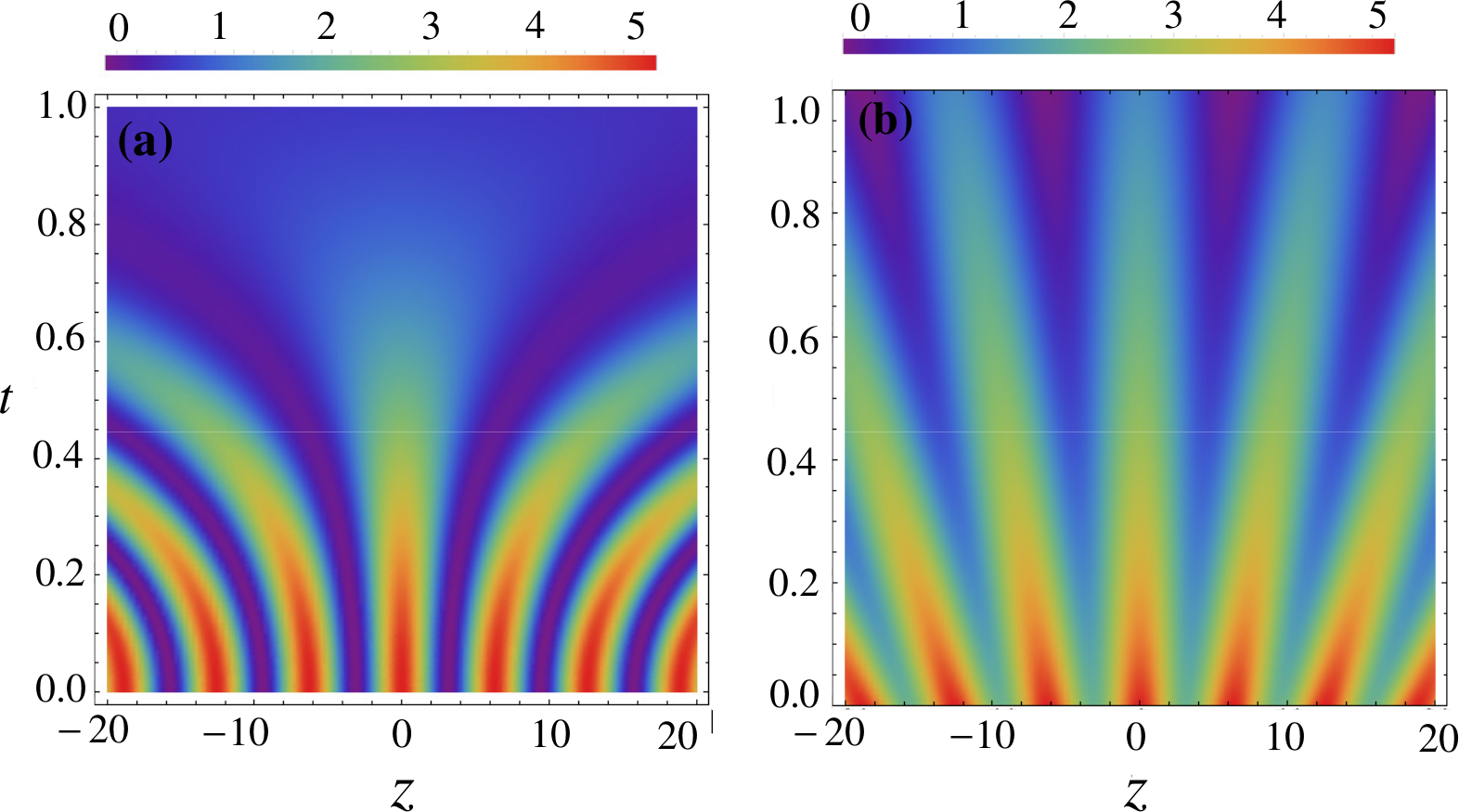}
\caption{Fig.(a) depicts the sinusoidal BEC profile in presence of an expulsive oscillator, where the chirped BEC
         undergoes expansion. The distortion in the density profile appears due to the presence of the trap.  Fig.(b) corresponds to a parameter domain when harmonic trap is switched off (for moving lattice), where BEC
         undergoes expansion, without any distortion.  The parameter values used here are same as in Fig.(\ref{density}).}
\label{densitycd}
\end{center}
\end{figure}

Here, we discuss the dynamics of BEC in an optical lattice potential, when the harmonic trap is switched off, where the condensate has no acceleration and hence moves with a constant velocity. The inverse of width and the amplitude of the lattice potential have been allowed to vary with time. In this case, the chirp parameter $c(t)$ can be determined by taking $M(t) = 0$ in Eq. (\ref{ricatti}): $\frac{\partial c(t)}{\partial t} = c^{2}(t)$. In this scenario, the solution retains its original form. The position of the matter wave can be found from the relation: $l(t) =  l_{0} (1 + c_{0} t)$, showing a constant velocity $v(t) = l_{0}c_{0} \equiv u$. It is interesting to note that in the absence of harmonic trap, the center of mass motion no longer oscillates, but follows a linear profile with time. In the case, with $c(t=0) = -1/c_{0}$, we find $c(t) = -1/(t + c_{0})$ and $k(t) = k_{0}c_{0}/(t + c_{0})$. As time increases, the inverse of width, as well as the amplitude of the matter wave decreases, leading to the spreading of the BEC, as depicted in Fig.(\ref{densitycd}b). In this case, since the harmonic trap is switched off, the density profile is smooth and no distortion has been seen.

\section{Energy of the excitation}

\subsection{Analytical results}
To explore the dynamics of this parametrically driven system, we study the temporal behavior of energy, both in the absence and presence of the trap. The energy of the condensate can be obtained from the following equation:
\begin{eqnarray}
E = \int dz \left[\frac{1}{2}\left|\frac{\partial \psi}{\partial z}\right|^{2} + (V_{l}(z,t) - \nu(t)) |\psi|^{2} + \frac{1}{2} g(t) |\psi|^{4} \right],
\label{energyeq}
\end{eqnarray}
\begin{figure}[t!]
\begin{center}
\includegraphics[scale=0.4]{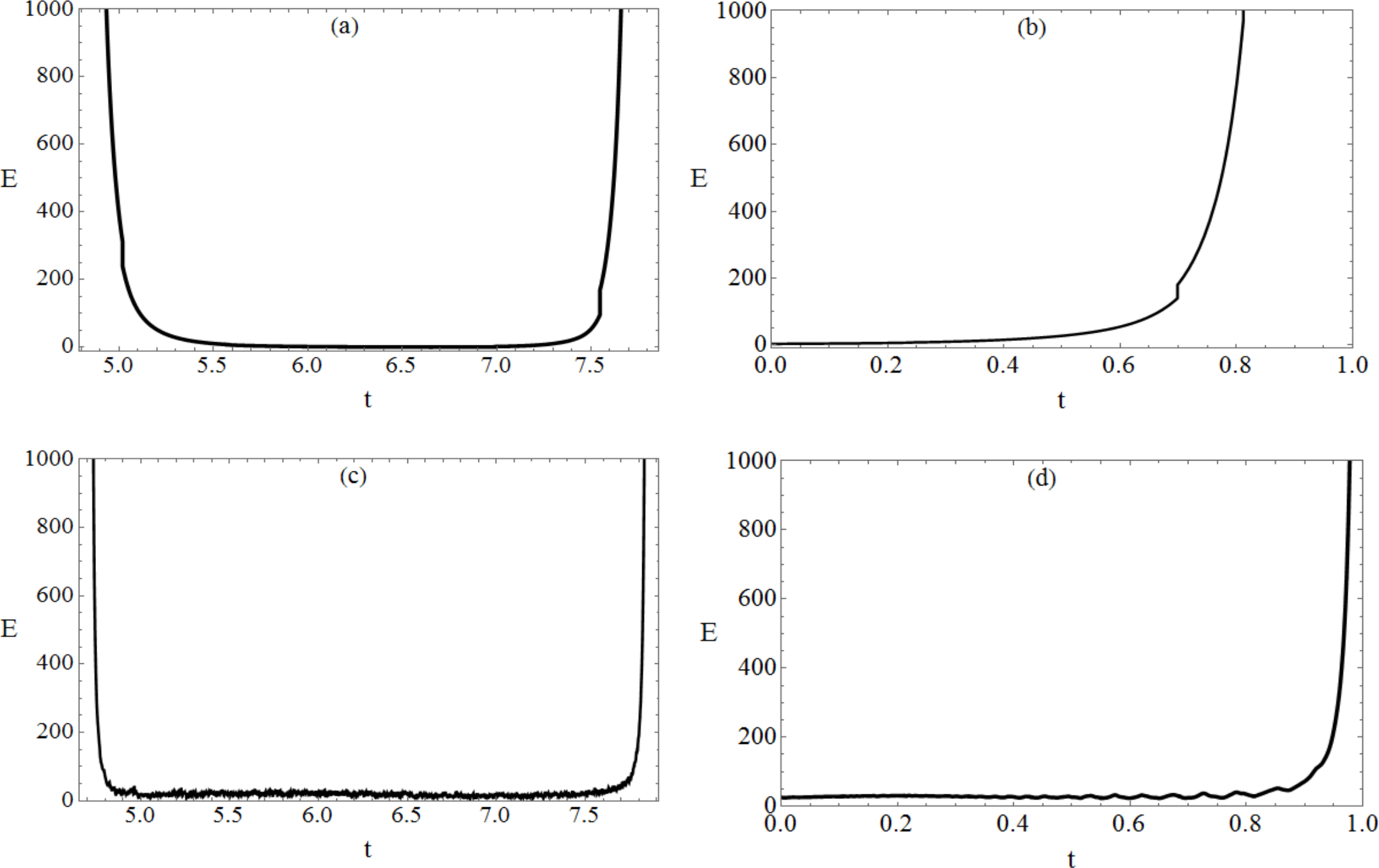}
\caption{Fig. (a) shows the nonlinear resonances in BEC in presence of both harmonic oscillator and lattice potential. Inset depicts the full energy profile, which shows the occurence of resonances periodically.  In Fig. (b), it is shown that resonance also occurs when the harmonic trap is switched off. This is akin to the nonlinear pulse compression, seen in \cite{moores1996}. Fig. (c) shows the nonlinear resonances obtained from numerical investigation, when both harmonic trap and lattice potential are present. Fig. (d) shows the numerical result of the nonlinear resonance, when trap is switched off. The energy has been scaled by the recoil energy $E_{R}$. The parameter values considered here are same as in Fig. (1). }
\label{energyw}
\end{center}
\end{figure}
with the terms in the right-hand side, respectively, representing its kinetic energy, the potential energy due to both harmonic and optical lattice along with a contribution from the chemical potential and interaction energy. The expression for the energy is too cumbersome to list and hence, we concentrate on the energy spectrum directly. A closer look at the energy spectrum in Fig.(\ref{energyw}), reveals that it actually mimics the density distributions, shown in Figs.(\ref{density}). As seen in Fig.(\ref{energyw}a), the condensate shows a rapid nonlinear resonant increase in energy in the presence of a harmonic trap and lattice potential. At certain values of the scaled time variable, it undergoes rapid nonlinear compression, which in turn mimics the occurrence of resonances in this system. These nonlinear resonances occur periodically at the point of nonlinear compression of BEC, where the density takes its maximum value. The contribution of the quadratic chirped phase to the kinetic energy is solely responsible for this phenomenon. In other words, the resonances occur when the driving frequency of the potential matches with the natural frequency of the system. The matter wave changes its direction at the point of nonlinear compression. This is akin to recently observed resonant behavior in an optical lattice, where energy transfer takes place between two bands \cite{fabbri2009}. It is worth mentioning that the nonlinear resonance also takes place when the harmonic trap is switched off. Fig.(\ref{energyw}b) shows an effective nonlinear compression when the harmonic trap is switched off. Rapid increase in energy is found with time for this driven system around $t \sim 0.8$. The thousand-fold increase in energy seen in both Figs.(\ref{energyw}a) and (\ref{energyw}b), is due to large nonlinear compression. This is analogous to the findings in \cite{moores1996}, in the context of effective pulse compression in nonlinear fiber optics.

\subsection{Numerical results}
Since, the main result of this paper is to show the generation of nonlinear resonances in a  BEC subject to an external OL potential, in this section we will only concentrate on the parameter regimes, where we observe the occurrence of the nonlinear resonance analytically. In our numerical simulation, we have used the split-step Crank-Nicholson (CN) method in order to solve the time dependent GP equation. In particular, we study the time evolution of GP equation and using Eq.(\ref{energyeq}), we compute the energy. In this simulation, we use the code {\it realtime1D.F}, published in \cite{muruganandam2009} and modify it as per the need of the present system. In our case, we choose $dz = 0.01$ and $dt = 0.00001$, which is well satisfied under the CN scheme: $dz^{2}/dt < 1$. We have introduced the nonlinearity $g(t)$ in NSTP $= 10000$ iterations. Due to the sinusoidal nature of the solutions obtained analytically, we consider a periodic function as an initial wavefunction, as well as periodic boundary conditions. The nonlinearity is then slowly introduced into the system in each time iteration until the desired value is attained. As a result, this leads to the final wavefunction, after NRUN $= 900000$ iterations \cite{muruganandam2009}. We then use Eq.(\ref{energyeq}) to compute the energy. To perform a systematic numerical study of Eq. (\ref{1D-NLSE}) and consequently Eq.(\ref{energyeq}), we use the same parameter values as provided in the analytical case. The numerical results are shown in Figs.(\ref{energyw}c) and (\ref{energyw}d). Fig.(\ref{energyw}c) depicts the energy as a function of time, when both harmonic trap and lattice potentials are present. It shows a rapid periodic increase in energy at certain time interval. This is exactly similar to that we observed in our analytical results, shown in Fig.(\ref{energyw}a). In the analytical case, as shown in Fig.(\ref{energyw}a), the resonances occur at $t \sim 5$ and $t \sim 7.6$, whereas, numerics shows the occurrence of the same around $t \sim 4.8$ and $t \sim 7.8$. In simulating the case for static OL, we use same NSTP to introduce the nonlinearity and consider NRUN $=100000$ for the convergence of the wavefunction. Fig.(\ref{energyw}d) shows the resonance increase in energy, when the harmonic trap is switched off. A sudden increase in energy is seen around $t \sim 0.9$. This is also similar to our analytical findings as shown in Fig.(\ref{energyw}b). In the analytical case, the resonance takes place around $t \sim 0.8$, whereas, numerics shows the same rapid increase in energy around $t \sim 0.9$. In both these cases, we observe a small difference at the point of the occurrence of nonlinear resonances, when comparing with our analytical results with numerics. This small difference occurs due to the assumption of the periodic boundary conditions in our simulations. The analytically obtained solutions through the self-similar method, do not follow the periodic boundary conditions. Therefore, we conclude that the spatial periodic boundary conditions are responsible for this small deviation. However, this does not alter the generation of nonlinear resonances, which is the main result of this paper. Therefore, we conclude that the results obtained from numerical simulation exactly support our analytical results.


\section{Dynamical superfluid-insulator transition}

We now examine the possibility of dynamical phase transition in this system.  In case of shallow lattice, a dynamical superfluid-insulator transition (DSIT), driven by modulational instability, was predicted by Smerzi et. al., using the mean-field discrete nonlinear Schr\"odinger equation \cite{smerzi2002dynamical}.  In the following year, Cataliotti et. al., experimentally observed this classical phase transition in the presence of a stationary optical lattice \cite{cataliotti2003superfluid}. It has been found in \cite{das2009loss} that, DSIT can also occur when both two and three-body interactions are present. Recently, Fallani et. al., have observed this dynamical instability in one dimensional moving optical lattice \cite{fallani2004}. Since the dynamical phase transition occurs at the point, where the energy of the system becomes non-analytic (NA), we have investigated the same from the exact energy expression for all the three cases, separately. As mentioned earlier, the energy expression is too long to be reported here, we will only concentrate on the non-analytic term, which in all the three cases is same and given by,
\begin{eqnarray}
E_{NA} = - \frac{4 k^{2}(t) (k^{2}(t) a (a + b) - \delta^{2})}{\sqrt{a (a + b)}} \tan^{-1}\frac{\sqrt{a} \tan (k(t) l(t))}{\sqrt{a + b}},
\end{eqnarray}

The energy is non-analytic at the following two points: $a + b = 0$ and $a = 0$, which respectively, lead to $2 \mu + 2 \alpha -1 = 0$ and $2 \mu - 1 = 0$. Both of these give $\delta = 0$, which reflects from Eq.(\ref{sc}) that the supercurrent is zero. For the density to be a finite positive quantity, we exclude the second point. Taking into account the first condition, the supercurrent vanishes at this point and the superfluid phase transits to an insulating phase, whose wave function is given by,
\begin{eqnarray}
\psi_{I}(z,t) = \sqrt{\frac{k(t) \alpha}{\kappa}} \sin{(k(t) z)} e^{i \phi(z,t)}.
\end{eqnarray}
Since, the superfluid phase is associated with a finite phase factor $\chi(T)$, the number of atoms on each lattice site are unknown. Therefore, the atoms are free to move within the lattice and can easily tunnel from one lattice site to another. The phase coherence in the superfluid phase can be experimentally observed through the formation of the interference pattern.  The modulational instability, occurring in the system, leads to a loss of phase coherence and hence the superfluid phase transits into the insulating phase. The insulating phase is not characterized by a definite phase and hence, the number of atoms per lattice site is fixed. Since the phase coherence is lost in the insulating phase, no interference pattern will be formed.


\section{Stability analysis: VK criterion}

For a positive semi-definite density of the sinusoidal excitation in NLS-type equations, the stability can be examined through the Vakhitov-Kolokolov criterion \cite{vakhitov1974}. It states that a necessary stability condition is a positive slope in the dependence of the chemical potential on the number of atoms per lattice site $N$; i.e., $\partial N/\partial \mu > 0$. If $\frac{\partial N}{\partial \mu} < 0$, the solutions are unstable and are marginally stable, if $\frac{\partial N}{\partial \mu} = 0$ \cite{weinstein1986,pelinovsky1996}. It is worth mentioning that, in some cases, e.g., for gap solitons obtained in \cite{malomed2010}, the VK criterion does not apply. The stability of the gap soliton families follow an ``anti-VK" criterion: $\partial \mu/\partial N > 0$. Therefore, in order to obtain the stability condition, we calculate the number of atoms per lattice site,
\begin{eqnarray}
N = \frac{k(t) (2 \mu - \alpha - 1)}{2 \kappa} - \frac{\alpha}{4 \pi \kappa} \sin (2 k(t) \pi),
\end{eqnarray}
which can be controlled through chirp management. We observe that for all the solutions, found in the previous section,
\begin{eqnarray}
\frac{\partial N}{\partial \mu} = \frac{k(t)}{\kappa},
\end{eqnarray}
showing that the stability of the obtained sinusoidal solutions depends on the chirp controlled inverse of the width of the condensate, as well as on the nature of the atom-atom interaction. Therefore, we conclude that the solutions are stable for repulsive atom-atom interaction ($\kappa > 0$).


\section{Conclusion}

In summary, we present a detailed study of the parametrically forced BEC in the presence of a general time modulated optical lattice and a harmonic trap, with both attractive and repulsive two-body interaction. The interaction between atoms and the lattice leads to sinusoidal excitations, which can be controlled by chirp management. We show through the density distribution and energy spectrum that BEC can be made to accelerate and undergo rapid nonlinear compression in the presence of a harmonic trap. A resonant increase in energy is observed at certain points of time, when the system couples to the harmonic trap. The combined effect of lattice and trap leads to the generation of these nonlinear resonances at the point, where the matter wave changes its direction. For the static case, when the harmonic trap is switched off, BEC undergoes a rapid nonlinear compression, analogous to an effective pulse compression in nonlinear fiber optics. Both these observations of resonances are well supported by numerical simulation. We also show that a classical dynamical phase transition occurs in the system, where the loss of superfluidity takes the superfluid phase to an insulating phase. The stability of the obtained solutions is investigated using well known VK criterion. We find that the solutions are stable for repulsive nonlinearity in the present case, though it strongly depends on the nature of the chirped phase. \\

\section*{Acknowledgement}
PD thanks Anirban Pathak for many useful discussions. PD acknowledges the financial support from T\"UB\.{I}TAK-1001,  Grant No. 114F170. PD also acknowledges Indian Institute of Science Education and Research Kolkata for providing all the facilities, where this work has been started.

\section*{References}

\bibliography{ms}
\bibliographystyle{iopart-num}

\end{document}